\UseRawInputEncoding

\documentclass[12pt]{article}
\usepackage{curves}
\usepackage{amsmath, amssymb, lamsarrow}
\usepackage{amssymb}

\usepackage{hyperref}
\def\mineappendix{
	\setcounter{section}{1}
	\setcounter{subsection}{0}
	\def\thesection{\Alph{section}}
	\def\sectionap{\@startsection  {section}{1}{\z@}
		{-3.5ex plus-1ex minus-.2ex} {0ex plus.2ex}
		{\reset@font\Large\bf  Appendix:  \, }
	}
}
\makeatother
\def\Proclaim #1. #2\par{\bigbreak\noindent{\sc#1.\enspace}{\it#2}\par}

\newtheorem{lem}{Lemma}[section]

\newtheorem{thm}[lem]{Theorem}
\newtheorem{pro}[lem]{Proposition}

\newcommand{\la}{\lambda}

\newenvironment{ack}{\noindent \textbf{Acknowledgments}.}

\title{Q-polynomial expansion for Br\'{e}zin-Gross-Witten tau-function}
\author{Xiaobo Liu \thanks{Research was partially supported by NSFC grants 11890662 and 11890660.}, Chenglang Yang}
\date{}

\begin{document}
\maketitle

\begin{abstract}
	In this paper, we prove a conjecture of Alexandrov that the generalized Br\'{e}zin-Gross-Witten tau-functions are hypergeometric
tau functions of BKP hierarchy after re-scaling. In particular, this shows that the original BGW tau-function, which has enumerative geometric interpretations, can be represented as a linear combination of Schur Q-polynomials with simple coefficients.
\end{abstract}

\section{Introduction}

The BGW tau-function, denoted by $\tau_{BGW}$, was introduced by Br\'{e}zin, Gross, and Witten  for studying lattice gauge theory in 1980 (c.f. \cite{BG} and \cite{GW}). Mironov, Morozov, and Semenoff showed that the BGW model can be considered as a particular case of the
generalized Kontsevich model and the partition function  $\tau_{BGW}$ is a tau function of the KdV hierarchy (c.f. \cite{MMS}).
In \cite{N17}, Norbury gave a conjectural enumerative geometric interpretations for $\tau_{BGW}$, i.e. it is
the generating function of intersection numbers of certain classes on the moduli spaces of stable curves (see Section \ref{sec:BGW} for precise definition). Hence  $\tau_{BGW}$ shares many similar properties with  Kontsevich-Witten tau function.
In \cite{MM}, Mironov and Morozov conjectured a simple expansion formula for Kontsevich-Witten tau function in terms of Schur's Q-polynomials. A proof of this conjecture using Virasoro constraints was recently given in \cite{LY}.
Inspired by Mironov-Morozov's conjecture, Alexandrov proposed similar conjectures for the BGW tau function and its generalizations
in \cite{Alex20}.
The main purpose of this paper is to prove Alexandrov's conjectures.

Let $Q_{\la}$ be the Schur Q-polynomial associated to a partition $\la$ (see Section \ref{sec:Q} for precise definition).
We will consider $Q_{\la}$ as a polynomial of variables $\textbf{t}:=(t_1, t_3, \cdots)$.
These polynomials are tau functions of BKP hierarchy (c.f. \cite{Y} and \cite{KL}).
In this paper
we will prove the following formula for BGW tau function which was conjectured by Alexandrov (i.e. Conjecture 1 in \cite{Alex20}):
\begin{thm}\label{thm:A=BGW}
	The Br\'{e}zin-Gross-Witten tau-function has the following expansion
	\begin{align*}
	\tau_{BGW}(\textbf{t})=\sum_{\lambda\in DP} 2^{-l(\la)} \bigg(\frac{\hbar}{16}\bigg)^{|\la|}
            \frac{Q_\lambda(\delta_{k,1})^3}{Q_{2\lambda}(\delta_{k,1})^2} \cdot Q_\lambda(\textbf{t}),
	\end{align*}
   where $DP$ is the set of all strict partitions, $l(\la)$ and $|\la|$ are the length and size of $\la$ respectively,  and  $\hbar$ is a formal parameter.
\end{thm}
In the above formula, $Q_\lambda(\delta_{k,1})$ is the value of $Q_{\la}$ at the point
$\textbf{t}=(1, 0, 0, \cdots)$. It is given by a simple formula \eqref{eqn:hook}
which is related to the hook length formula.

The generalized BGW model was introduced by Mironov, Morozov, and Semenoff in \cite{MMS}.
This is a family of matrix models indexed by a complex number $N$.
When $N=0$, it is the original BGW model.
The partition functions of these models, denoted by $\tau_{BGW}^{(N)}$, are also tau functions
of the KdV hierarchy and satisfy the Virasoro constraints. A description of $\tau_{BGW}^{(N)}$
using cut-and-join operators was given by Alexandrov in \cite{Alex18}.
These tau functions can be considered as a deformation of $\tau_{BGW}$
analogous to the Kontsevich-Penner deformation
of the Kontsevich-Witten tau function.
In this paper, we will prove the following formula which was also conjectured by  Alexandrov
(i.e. Conjecture 2 in \cite{Alex20}):
\begin{thm}\label{thm:A=gBGW}
	The generalized Br\'{e}zin-Gross-Witten tau-function has the following expansion
	\begin{equation}\label{eqn:A=gBGW}
	\tau_{BGW}^{(N)}(\textbf{t})=\sum_{\lambda\in DP} \bigg( \frac{\hbar}{16} \bigg)^{|\la|}2^{-l(\la)}\theta_\la
        Q_\lambda(\delta_{k,1})Q_\lambda(\textbf{t}),
	\end{equation}
	where
    \begin{equation}  \label{eqn:thetala}
           \theta_\la :=\prod_{j=1}^{l} \prod_{k=1}^{\la_j} \theta(k)
    \end{equation}
    for $\la=(\la_1, \cdots, \la_l)$ and
    \begin{equation} \label{eqn:theta}
        \theta(z):=(2z-1)^2-4N^2.
    \end{equation}
\end{thm}
In particular, this formula implies that $\tau_{BGW}^{(N)}(\textbf{t}/2)$  are  hypergeometric tau functions of BKP hierarchy
as defined by Orlov in \cite{Or}. When $N=0$, this theorem is equivalent to Theorem \ref{thm:A=BGW} due to equation
\eqref{eqn:hook}.

This paper is organized as follows. In section \ref{sec:pre}, we review definitions and basic properties of generalized BGW tau-functions and Schur Q-polynomials. In section \ref{sec:Vira}, we prove that the right hand side of equation \eqref{eqn:A=gBGW} satisfies the
Virasoro constraints of $\tau_{BGW}^{(N)}(\textbf{t})$. Since the Virasoro constraints uniquely fix the tau function
up to a constrant (c.f. \cite{Alex18}), this gives a proof of
Theorems \ref{thm:A=gBGW} and \ref{thm:A=BGW}.

The result of this paper was forecasted in \cite{LY} which focused on the Kontsevich-Witten tau function. Three days before the current paper was posted on the arXiv (c.f. arXiv:2104.01357), a new preprint \cite{Alex21} appeared which also gives a proof of Theorems  \ref{thm:A=BGW} and \ref{thm:A=gBGW} via a completely different approach using Boson-Fermion correspondence.

\begin{ack}
The authors would like to thank  Leonid Chekhov and Alexei Morozov for their interests in this work.
\end{ack}

\section{Preliminaries}\label{sec:pre}

\subsection{Br\'{e}zin-Gross-Witten tau-function and its generalizations}
\label{sec:BGW}

The BGW model was originally proposed in \cite{BG} and \cite{GW} as a unitary matrix model.
It was conjectured in \cite{N17} that the partition function of this model, i.e. $\tau_{BGW}$,
has the following geometric interpretation.

Let $\overline{\mathcal{M}}_{g,n}$ be the moduli space of stable genus $g$ curves $C$ with $n$ distinct
smooth marked points $x_1, \ldots, x_n \in C$.
For each $i \in \{1, \cdots, n\}$, there is a tautological class $\psi_i \in H^2(\overline{\mathcal{M}}_{g,n})$,
which is the first Chern class of the line bundle on $\overline{\mathcal{M}}_{g,n}$ whose
fiber at a point $(C; x_1, \cdots, x_n) \in \overline{\mathcal{M}}_{g,n}$ is the cotangent
space of $C$ at the marked point $x_i$.
 The Kontsevich-Witten tau function is the generating function of intersection numbers of these
$\psi$-classes  (c.f. \cite{W} and \cite{K}).
To recover the BGW tau function, Norbury constructed a new family of classes
$\Theta_{g,n}\in H^{4g-4+2n}(\overline{\mathcal{M}}_{g,n})$, which are well behaved with respect to
pullbacks of gluing and forgetful maps among moduli spaces of stable curves. 
Following notations in \cite{Alex20}, define intersection numbers
\[\langle \tau_{k_1} \tau_{k_2} \cdots \tau_{k_n} \rangle^{\Theta}_g := \, \int_{\overline{\mathcal{M}}_{g,n}} \Theta_{g,n} \, \psi_1^{k_1} \psi_2^{k_2} \cdots \psi_n^{k_n} \]
for non-negative integers $k_1, k_2, \ldots, k_n$. Since the complex dimension of $\overline{\mathcal{M}}_{g,n}$ is
$3g-3+n$, these intersection numbers are non-zero only if
$\sum_{i=1}^n k_i = g-1$.

Let
\[ F_{g,n}^{\Theta}(\textbf{t}):= \frac{1}{n!} \sum_{k_1, \cdots, k_n \geq 0} \langle \tau_{k_1} \cdots \tau_{k_n} \rangle^{\Theta}_g
      \prod_{i=1}^n (2k_i+1)!! t_{2k_i +1}.\]
Assign degree of $t_{k}$ to be $k$ for all $k$. Then $F_{g,n}^{\Theta}(\textbf{t})$ is a homogeneous polynomial
of degree $2g-2+n$ by the above dimension constraint. Norbury's conjecture (stated as a theorem in the first three versions of \cite{N17}) can be stated as
\[ \tau_{BGW}(\textbf{t}) = \exp \bigg( \sum_{g=0}^{\infty} \sum_{n=0}^{\infty} \hbar^{2g-2+n} F_{g,n}^{\Theta}(\textbf{t}) \bigg).
\]
This conjecture  has been verified up to genus $7$. 
The Virasoro constraints for $\tau_{BGW}$  were obtained in \cite{GN}.

Like BGW model, the generalized BGW model proposed in \cite{MMS} is a family of matrix models
indexed by a complex number $N$. The partition function $\tau_{BGW}^{(N)}(\textbf{t})$ of this model
is a tau function of KdV heirarchy for all $N$. When $N=0$,
\[ \tau_{BGW}^{(0)}=\tau_{BGW} \]
 is the original BGW tau function explained above.
However, for $N \neq 0$, enumerative geometric interpretation of   $\tau_{BGW}^{(N)}$
is not known at the moment, although existence of such an interpretation is expected (see, for example, \cite{Alex20}).
It was shown in \cite{AC} that a similar one-parameter deformation of BGW model can be transformed to the Kontsevich model after a shift of times except for the original BGW model.
On the other hand,
 Alexandrov proved in \cite{Alex18} that for any $N$, the generalized BGW tau function $\tau_{BGW}^{(N)}(\textbf{t})$
is uniquely determined by the {\it normalization condition}
\begin{equation} \label{eqn:normalization}
 \tau_{BGW}^{(N)} (0) = 1
\end{equation}
and the {\it Virasoro constraints}
\begin{equation} \label{eqn:VirConst}
 L_m^{(N)} \, \tau_{BGW}^{(N)}=0 {\rm \,\,\, \,\,\, for \,\,\,\,\,\,} m \geq 0,
\end{equation}
where
\begin{align*}
L_{m}^{(N)}:=\frac{1}{4}\sum_{a+b=2m}\frac{\partial^2}{\partial t_a \partial t_b}
+\frac{1}{2}\sum_{k \geq 1 \atop k \,\, odd}kt_k\frac{\partial}{\partial t_{k+2m}}
-\frac{1}{2\hbar}\frac{\partial}{\partial t_{2m+1}}
+\frac{1-4N^2}{16}\delta_{m,0}.
\end{align*}
In this paper, we will take this property as the definition of $\tau_{BGW}^{(N)}$.

The operators $L_{m}^{(N)}$ satisfy the bracket relation
\[[ L_k^{(N)}, L_l^{(N)}]=(k-l) L_{k+l}^{(N)} \]
for all $k, l \geq 0$.
So they form a half branch of the Virasoro algebra. In particular, the first three operators $ L_{0}^{(N)}$,  $L_1^{(N)}$ and $ L_2^{(N)}$
generate all other operators $L_{k}^{(N)}$ for $k>2$.
Thus, to prove a function satisfying the Virasoro constraints \eqref{eqn:VirConst}, we just need to prove that it satisfies the $ L_{k}^{(N)}$-constraint for $k=0,1,2$. Since $L_{k}^{(N)}$ does not depend on $N$ for $k>0$, we will simply write it as $L_{k}$.
More explicitly, the first three Virasoro operators are given by
\begin{align} \label{eqn:Vira}
{L}_0^{(N)}=& \frac{1}{2}\sum_{k \geq 1 \atop k \, odd} k t_k \frac{\partial}{\partial t_{k}}
                   -\frac{1}{2\hbar}\frac{\partial}{\partial t_{1}}+\frac{1-4N^2}{16},
                      \nonumber \\
{L}_1 = & \frac{1}{2}\sum_{k \geq 1 \atop k \, odd}kt_{k}\frac{\partial}{\partial t_{k+2}}
            +\frac{1}{4} \frac{\partial^2}{\partial t_{1}\partial t_{1}}-\frac{1}{2\hbar}\frac{\partial}{\partial t_{3}}, \nonumber \\
{L}_2 =& \frac{1}{2}\sum_{k \geq 1 \atop k \, odd}kt_{k}\frac{\partial}{\partial t_{k+4}}
          +\frac{1}{2}\frac{\partial^2}{\partial t_{1}\partial t_{3}}-\frac{1}{2\hbar}\frac{\partial}{\partial t_{5}}.
\end{align}

\subsection{Schur Q-polynomial}
\label{sec:Q}

In this paper, we will follow basic conventions for partitions and Q-polynomials in \cite{LY}.
A partition of length $l(\la)=l$ is a sequence of non-negative integers
$\la=(\la_1, \cdots, \la_l)$. The size of $\la$ is defined to be
$|\la|:=\sum_{i=1}^l \la_i$.
A partition $\la$ is {\it positive} if all its parts $\la_i$ are positive.
It is {\it weakly positive} if it has at most one part equal to $0$.
It is {\it strict} if $\la_1 > \la_2 > \cdots > \la_l >0$.
In particular, a strict partition is always positive.
The set of all strict partitions is denoted by $DP$.
For convenience, the {\it empty partition} $\emptyset$, i.e. the partition which has no parts,
is considered to be a strict partition.
We will also use the following notations: For any integers $1 \leq i_1 < \cdots < i_n \leq l$,
\begin{equation}
\lambda^{\{i_1, \cdots, i_n\}}:=(\lambda_1, \cdots, \widehat{\lambda}_{i_1}, \cdots, \widehat{\lambda}_{i_n}, \cdots, \lambda_l),
\end{equation}
where $\widehat{\lambda}_{i}$ means that the $i$-th part should be deleted from the partition.
If $\la$ is a partition and $a_1, \cdots, a_k$ are non-negative integers, then
\[ (\lambda, a_1, \cdots, a_k):=(\lambda_1, \cdots, \lambda_l, a_1, \cdots, a_k). \]

Q-polynomials were introduced by Schur in the study of projective representations of the symmetric group.
As explained in Macdonald's book \cite{Mac}, there are several equivalent definitions for
such polynomials. For most of time, Schur's original definition using Pfaffian would be sufficient for us.
It starts with a sequence of polynomials $q_k(\mathbf{t})$ defined by
\begin{equation} \label{eqn:Ql1}
\sum\limits_{k=0}^\infty q_k(\mathbf{t})z^k=\exp{ \bigg( 2\sum_{k=0}^{\infty} t_{2k+1} \, z^{2k+1} \bigg)},
\end{equation}
where $z$ is a formal parameter. For a pair of non-negative integers $(r,s) \neq (0,0)$,
define
\begin{equation} \label{eqn:Ql2}
A_{r,s}:=q_r(\mathbf{t})q_s(\mathbf{t})+2\sum_{i=1}^s(-1)^iq_{r+i}(\mathbf{t})q_{s-i}(\mathbf{t}),
\end{equation}
and set $A_{0,0}:=0$. It turns out that $A_{r,s}$ is skew symmetric with respect to $r$ and $s$.
If $\lambda=(\lambda_1,...,\lambda_{2m})$ is a weakly positive partition of even length, we can
define the associated {\it Schur Q-polynomial}
 by the Pfaffian:
\begin{equation} \label{eqn:Qwp}
Q_\lambda(\textbf{t}):=\text{Pf} \big( A_{\lambda_i,\lambda_j} \big)_{1\leq i,j\leq 2m}.
\end{equation}
By properties of Pfaffian, $Q_\lambda$ is skew symmetric with respect to permutations of parts of $\la$.
In particular $Q_{\la}=0$ if $\la$ has two equal positive parts.

In \cite{Mac} pages 262-263, a larger system of functions $Q_\lambda$ were defined
for all $\la=(\la_1, \cdots, \la_l) \in \mathbb{Z}^l$. In this paper, we only need a portion of such functions,
i.e. $Q_{\la}$ for $\la$ with at most one  negative part.
These functions are uniquely determined by the following rules:
\begin{itemize}
    \item If $\la$ is weakly positive of even length, it is given by equation \eqref{eqn:Qwp}.
    \item $Q_{(\la,0)}= Q_{\la}$ for all $\la$.
	\item If for some $i<l(\la)$, a partition $\tilde{\la}$ is obtained from $\la$ by switching $\la_i$ and $\la_{i+1}$ which are not both equal to $0$, then $Q_{\tilde{\la}} = - Q_{\la}$.
    \item Assume $\la$ has exactly one part $\la_i<0$ and  $\la_j \geq 0$ for all $j \neq i$.
    If there exists $j>i$ such that $\la_j = -\la_{i}$ and $\la_k \neq - \la_i$ for all $k>i$ and $k \neq j$, then define
	\begin{equation} \label{eqn:Q-}
         Q_{\lambda} := (-1)^{j-i-1+\lambda_j} \, 2 \,Q_{\la^{\{i,j\}}}.
    \end{equation}
	Otherwise define $Q_{\lambda}:=0$.
    \item $Q_{\emptyset}=1$.
\end{itemize}
Note that the notion of Q-polynomial associated with $\la$ used in \cite{MM} and \cite{Alex20} is equal to $2^{-l(\la)/2} Q_{\la}$.
So formulas in this paper differ from corresponding formulas in \cite{MM} and \cite{Alex20}  by a suitable factor.

Assign the degree of $t_k$ to be $k$ for all $k$. Then $Q_\lambda(\textbf{t})$ is a homogeneous polynomial of degree $|\lambda|$.
Moreover $\{Q_\lambda(\textbf{t}) \mid \lambda\in DP \}$ form a
 basis of $\mathbb{Q}[t_1,t_3,...]$, which has
a standard inner product such that
\begin{align}\label{eqn:inner prod}
\langle Q_\lambda,Q_\mu\rangle=2^{l(\lambda)}\delta_{\lambda,\mu} \text{\,\,\, if\ } \lambda, \mu \in DP.
\end{align}
For any operator $f$ on $\mathbb{Q}[t_1,t_3,...]$,  the adjoint operator $f^{\perp}$ is defined by
\[ \langle f^{\perp} p_1(\textbf{t}), p_2(\textbf{t}) \rangle = \langle p_1(\textbf{t}), f p_2(\textbf{t}) \rangle \]
for all $p_1, p_2 \in \mathbb{Q}[t_1,t_3,...]$.
Each polynomial $p \in  \mathbb{Q}[t_1,t_3,...]$ can be considered as an operator which
acts on $\mathbb{Q}[t_1,t_3,...]$ by multiplication.

Let $r$ be a positive odd integer. We then have
\begin{align}\label{eqn:ad-relation}
t_r^\perp=\frac{1}{2r}\frac{\partial}{\partial t_r},
\end{align}
\begin{equation}\label{eqn:dQ}
\frac{1}{2}\frac{\partial}{\partial t_r} Q_\lambda=\sum_{i=1}^{l(\lambda)}Q_{\lambda-r\epsilon_i},
\end{equation}
and
\begin{align}\label{eqn:multi pr}
	rt_rQ_\lambda=\sum_{i=1}^{l(\lambda)}Q_{\lambda+r\epsilon_i} +\frac{1}{2}Q_{(\lambda,r)}
	+\sum_{k=1}^{r-1}\frac{(-1)^{r-k}}{4}Q_{(\lambda,k,r-k)}
	\end{align}
for any partition $\la$, where
\[ {\lambda \pm r\epsilon_i}:=(\lambda_1,...,\lambda_i \pm r,...,\lambda_l)\]
(see, for example,  Lemma 2.2 and Corollary 2.3 in \cite{LY}, and \cite{Mac} page 266).

The following formulas were proved by Aokage, Shinkawa, and Yamada in \cite{ASY}: For all strict partitions $\la$,
\begin{align}\label{eqn:L'-formula}
L_1'Q_{\lambda}(\textbf{t})=\sum_{i=1}^{l(\lambda)} (\lambda_i-1) Q_{\lambda-2\epsilon_i}(\textbf{t})
\hspace{15pt} {\rm and} \hspace{15pt}
L_2'Q_{\lambda}(\textbf{t})=\sum_{i=1}^{l(\lambda)} (\lambda_i-2) Q_{\lambda-4\epsilon_i}(\textbf{t}),
\end{align}
where
\begin{align}
L_1':=\sum_{k \geq 1 \atop k \, odd}kt_k\frac{\partial}{\partial t_{k+2}}+\frac{1}{4}\frac{\partial^2}{\partial t_{1}\partial t_{1}}
\hspace{15pt} {\rm and} \hspace{15pt}
L_2' :=\sum_{k \geq 1 \atop k \, odd}kt_k\frac{\partial}{\partial t_{k+4}}+\frac{1}{2}\frac{\partial^2}{\partial t_{1}\partial t_{3}}.
\label{eqn:L2'}
\end{align}
Note that iterated brackets of $L_1'$ and $L_2'$ also generate a half branch of Virasoro algebra. These Virasoro operators
are different from the Virasoro operators for generalized BGW tau functions given by equation \eqref{eqn:Vira}.
Since the definition of Q-polynomials in \cite{ASY} is slightly different from the definition in this paper, we have modified
the coefficients of $L_1'$ and $L_2'$ to accommodate such difference (see \cite{LY} Section 2.2 for more explanations).

\subsection{Properties of $Q_\la(\delta_{k,1})$}
\label{sec:B}

For any partition $\la$,  define
\[B_\la:=Q_\la(\delta_{k,1}).\]
It is well known that
\begin{align}\label{eqn:hook}
B_\la=\frac{2^{|\la|}}{\la!}\prod_{i<j}\frac{\la_i-\la_j}{\la_i+\la_j}
\end{align}
for weakly positive partitions $\la$, where $\la ! := \prod_{i=1}^{l(\la)} \la_i !$
  (see, for example, equation (3.3) in \cite{Alex20}).
This formula is related to the hook length formula.
For most cases, we only need this formula for $l(\la)=1$ or $l(\la)=2$. They are given by
\begin{align}\label{eqn:hook1}
B_{(k)} = \frac{2^{k}}{k!}
\end{align}
and
\begin{align}\label{eqn:hook2}
B_{(k,m)}= \frac{2^{k+m}}{k!m!} \, \frac{k-m}{k+m}
\end{align}
for all non-negative integers $k$ and $m$ which are not both equal to $0$.
Equation \eqref{eqn:hook} implies
\begin{align}\label{eqn:Q/Q}
\frac{B_\la}{B_{2\la}}=\prod_{j=1}^{l(\la)}(2\la_j-1)!!
\end{align}
for all weakly positive partitions $\la$.

Since for any weakly positive partition $\la$ with even length, $B_{\la}$
is given by the Pfaffian of a skew symmetric matrix, it satisfies the standard
recursion relation for Pfaffian
\begin{align}\label{eqn:BRec2}
B_\la=\sum_{i=2}^{l(\la)} (-1)^i B_{(\la_1,\la_i)}B_{\la^{\{1,i\}}}
\end{align}
(c.f. Equation (2.4) in \cite{O} and Theorem 9.14 in \cite{HH}).
By skew symmetry,  we can expand $B_{\la}$ with respect to any part $\la_j$ to obtain
\begin{align}\label{eqn:BRec2j}
B_\la=(-1)^{j-1}\sum_{i=1 \atop i\neq j}^{l(\la)} (-1)^{\tilde{i}(j)} B_{(\la_j,\la_i)}B_{\la^{\{j,i\}}},
\end{align}
where
\begin{equation} \label{eqn:itilde}
\tilde{i}(j) := i-\delta_{i<j}
\end{equation}
with $\delta_{i<j}$  equal to $1$ if $i<j$ and equal to $0$ otherwise.

If $\la$ is a positive partition with odd length, then we can still use equations \eqref{eqn:BRec2} and \eqref{eqn:BRec2j} after replacing $\la$ by $(\la,0)$. In particular, applying equation \eqref{eqn:BRec2j} with $j=l(\la)+1$ for $(\la,0)$, we obtain
\begin{equation} \label{eqn:BRecOdd}
B_{\la}=\sum_{i=1}^{l(\la)} (-1)^{i+1} B_{(\la_i)} \, B_{\la^{\{i\}}}
\end{equation}
for any positive partition $\la$ with odd length.

The same proof for Lemma 3.6 in \cite{LY} shows the following
\begin{lem}\label{lem:even-1}
	If $\la$ is a positive partition with even length $l$,
	\[\sum_{i=1}^{l(\la)} (-1)^i B_{(\la_i)} \, B_{\la^{\{i\}}} =0.\]
\end{lem}
The same proof for Lemma 3.7 in \cite{LY} shows the following
\begin{lem}\label{lem:odd0}
 Let $\mu=(\mu_1, \cdots, \mu_l)$ be a weakly positive partition of odd length.
 Then equation \eqref{eqn:BRec2j} holds for $\la=(\mu, 0)$ with $j=1, \cdots, l$.
\end{lem}

\section{Proof of main theorems}
\label{sec:Vira}

Let $\tau_{N}$ be the function given by the right hand side of equation \eqref{eqn:A=gBGW}, i.e.
\begin{equation}
\tau_{N}(\textbf{t}):=\sum_{\lambda\in DP} \left( \frac{\hbar}{16} \right)^{|\la|} 2^{-l(\la)} \theta_\la B_\lambda Q_\lambda(\textbf{t})
\end{equation}
where $\theta_\la$ is defined by equations \eqref{eqn:thetala} and  \eqref{eqn:theta}, and $B_{\la}=Q_\lambda(\delta_{k,1})$. In particular,
$\theta_\la$ depends on $N$.
Alexandrov's conjecture for generalized BGW tau functions  can be restated as $\tau_{BGW}^{(N)}=\tau_{N}$
for all complex numbers $N$.

In this section, we will show that $\tau_{N}$  satisfies the Virasoro constraints
\begin{align} \label{eqn:VirA}
L_{m}^{(N)}\tau_{N}=0
\end{align}
for $m \geq 0$. Since $\tau_{BGW}^{(N)}$ satisfies the same Virasoro constraints and the same normalization condition, this will imply
Theorem \ref{thm:A=gBGW}.

Since $\{ Q_\la \mid \la \in DP\}$ form an orthogonal basis of $\mathbb{Q}[t_1, t_3, \cdots]$, equation \eqref{eqn:VirA}
is equivalent to
\[ \left\langle L_{m}^{(N)}\tau_{N}, \,\, Q_\la \right\rangle
= \left\langle \tau_{N},  \,\, \left(L_{m}^{(N)}\right)^\perp Q_\la \right\rangle = 0 \]
for all $\la \in DP$.
Therefore instead of computing the action of $L_{m}^{(N)}$ on $\tau_{N}$, we will
compute action of the adjoint operator $\left(L_{m}^{(N)}\right)^\perp$ on each  $Q_\la$ with $\la \in DP$.

Due to the  Virasoro bracket relation, it suffices to prove equation \eqref{eqn:VirA} for $m=0, 1, 2$.

\subsection{Action of Virasoro operators}
	
Recall operators $L_0^{(N)}, L_1, L_2$ are defined by equation \eqref{eqn:Vira}, and
operators $L_1', L_2'$ are defined by equation \eqref{eqn:L2'}. By equation \eqref{eqn:ad-relation}, we have
\begin{lem}\label{lem:LNperp}
\begin{align*}
(L_0^{(N)})^\perp=&\frac{1}{2}\sum_{k \geq 1 \atop k \,\, odd}kt_k\frac{\partial}{\partial t_{k}}+\frac{1-4N^2}{16}-\frac{t_1}{\hbar},\\
(L_1)^\perp=&\frac{1}{2}(L_1')^\perp+\frac{t_1t_1}{2}-\frac{3t_3}{\hbar},\\
(L_2)^\perp=&\frac{1}{2}(L_2')^\perp+3t_1t_3-\frac{5t_5}{\hbar}.
\end{align*}
\end{lem}

Note that the action of $(L_0^{(N)})^\perp$ is easy to compute since
\[ \sum_{k \geq 1 \atop k \,\, odd} k t_k \frac{\partial}{\partial t_{k}} Q_{\la} = |\la| Q_\la, \]
which follows from the fact that $Q_\la$ is homogeneous of degree $|\la|$. Multiplications by $t_1$, $t_3$, $t_5$ are given by equation
 \eqref{eqn:multi pr}.
 We also need to compute actions of other operators in the right hand sides
of equations in Lemma \ref{lem:LNperp}.
We start with computing  actions of $(L_1')^\perp$ and $(L_2')^\perp$ first.

\begin{lem}\label{lem:L12'adj}
For any strict partition $\la=(\la_1,...,\la_l)$,
\begin{align*}
(L_1')^\perp\cdot Q_\la=\sum_{i=1}^{l(\lambda)}(\lambda_i+1)Q_{\la+2\epsilon_i} + \frac{1}{2} Q_{(\la,2)},\\
(L_2')^\perp\cdot Q_\la=\sum_{i=1}^{l(\lambda)}(\lambda_i+2)Q_{\la+4\epsilon_i}+Q_{(\la,4)}-\frac{1}{2}Q_{(\la,3,1)}.
\end{align*}
\end{lem}
{\bf Proof}:
The formula for the action of $(L_2')^\perp$  has been proved in \cite{LY} Lemma 4.3. We only need to prove the first formula here.

Since $\{Q_\mu \mid \mu \in DP \}$ form an orthogonal basis of $\mathbb{Q}[t_1, t_3, \cdots]$, we have
\begin{equation} \label{eqn:L1'adjin}
(L_1')^\perp \cdot Q_\lambda
= \sum_{\mu \in {\rm DP}} 2^{-l(\mu)} \, \langle (L_1')^\perp \cdot Q_\lambda, \,\, Q_{\mu} \rangle \,\, Q_{\mu}.
\end{equation}
By skew symmetry of Q-polynomials, each summand on the right hand side of this equation is symmetric with
respect to permutations of $\mu$. Hence we can replace each $\mu$ in the above equation by any permutation of $\mu$.
After permutation, $\mu$ is still a partition with positive distinct parts.

By equation \eqref{eqn:L'-formula}, for $\mu = (\mu_1, \cdots, \mu_{l(\mu)})$,
\begin{equation} \label{eqn:L1'ml}
\langle (L_1')^\perp \cdot Q_\lambda, \,\, Q_{\mu} \rangle = \langle Q_\lambda, \,\, L_1' \cdot Q_{\mu} \rangle
=\sum_{j=1}^{l(\mu)} (\mu_j - 1) \langle Q_\lambda, \,\,   Q_{\mu - 2 \epsilon_j} \rangle.
\end{equation}
For this inner product to be non-zero, $\mu$ must have one of the following two forms.

{\bf Case (1)}, $\mu$ is a permutation of $\la + 2  \epsilon_i$ for some $i$ between $1$ and $l(\la)$. We may assume
$\mu = \la + 2  \epsilon_i$. By equation \eqref{eqn:L1'ml},
\[ \langle (L_1')^\perp \cdot Q_\lambda, \,\, Q_{\mu} \rangle
=\sum_{j=1}^{l(\la)} (\la_j + 2 \delta_{j,i} - 1) \langle Q_\lambda, \,\,   Q_{\la + 2  \epsilon_i - 2 \epsilon_j} \rangle.
\]
For $\langle Q_\lambda, \,\,   Q_{\la + 2 \epsilon_i - 2 \epsilon_j} \rangle \neq 0$,
$\la + 2  \epsilon_i - 2 \epsilon_j$ must be a permutation of $\la$. If $j \neq i$, this implies $\la_i + 2 = \la_j$ which is not possible
since $\mu = \la + 2  \epsilon_i$ must have distinct parts. Therefore we must have $j=i$ and
\begin{equation} \label{eqn:L1'c1}
\langle (L_1')^\perp \cdot Q_\lambda, \,\, Q_{\mu} \rangle
=(\la_i + 1) \langle Q_\lambda, \,\,   Q_{\la} \rangle = 2^{l(\la)} (\la_i+1)= 2^{l(\mu)} (\la_i+1)
\end{equation}
for $\mu = \la + 2  \epsilon_i$.

{\bf Case (2)}, $\mu$ is a permutation of $(\la, 2)$. We may assume
$\mu = (\la, 2)$. Since $\mu$ must have distinct parts,  $\la$ can not have parts equal to $2$.
By equation \eqref{eqn:L1'ml},
\[ \langle (L_1')^\perp \cdot Q_\lambda, \,\, Q_{\mu} \rangle
=\sum_{j=1}^{l(\la)} (\la_j  - 1) \langle Q_\lambda, \,\,   Q_{(\la-2\epsilon_j, 2)} \rangle
+ (2  - 1) \langle Q_\lambda, \,\,   Q_{(\la, 0)} \rangle.
\]
Note that $l(\la-2\epsilon_j, 2) = l(\la)+1$.  Since $\la$ does not have parts equal to $2$,
$\langle Q_\lambda, \,\,   Q_{(\la-2\epsilon_j, 2)} \rangle = 0$ for $1 \leq j \leq l(\la)$.
Moreover $Q_{(\la, 0)}=Q_{\la}$. So we have
\begin{equation} \label{eqn:L1'c2}
\langle (L_1')^\perp \cdot Q_\lambda, \,\, Q_{\mu} \rangle
=  \langle Q_\lambda, \,\,   Q_{\la} \rangle = 2^{l(\la)}= 2^{l(\mu)-1}
\end{equation}
for $\mu = (\la, 2)$.

Combining equations \eqref{eqn:L1'adjin}, \eqref{eqn:L1'c1}, \eqref{eqn:L1'c2},
we obtain the desired formula. The lemma is thus proved.
$\Box$
	
By repeatedly applying equation \eqref{eqn:multi pr}, we have
\begin{lem}\label{lem:t11t13}
\begin{align*}
t_1 t_1 Q_\la(\mathbf{t})=& \sum_{i,j=1}^{l(\la)} Q_{\la+\epsilon_i+\epsilon_j}(\textbf{t})
    +\sum_{i=1}^{l(\la)} Q_{(\la+\epsilon_i,1)}
    + \frac{1}{2} Q_{(\la,2)},\\
3t_1t_3Q_\la(\mathbf{t})=&\sum_{i,j=1}^{l(\lambda)} Q_{\lambda+\epsilon_i+3\epsilon_j}
+\frac{1}{2}\sum_{i=1}^{l(\la)} Q_{(\la+3\epsilon_i,1)}+\frac{1}{2} Q_{(\la,4)} \nonumber\\
+&\frac{1}{2}\sum_{i=1}^{l(\la)}Q_{(\la+\epsilon_i,3)}
-\frac{1}{2}\sum_{i=1}^{l(\la)}Q_{(\la+\epsilon_i,2,1)}
-\frac{1}{4}Q_{(\la,3,1)}.
\end{align*}
\end{lem}

Combining  above results from Lemmas \ref{lem:LNperp}, \ref{lem:L12'adj}, \ref{lem:t11t13}, and applying equation \eqref{eqn:multi pr}, we obtain
\begin{pro}\label{pro:DhL}
For any strict partition $\lambda=(\lambda_1,...,\lambda_{l})$, we have
{\allowdisplaybreaks
\begin{align}
( L_{0}^{(N)})^\perp\cdot Q_{\lambda}
=&  \left( \frac{|\la|}{2}+\frac{1-4N^2}{16} \right) Q_{\la}
    -\frac{1}{\hbar} \bigg( \sum_{i=1}^{l(\la)}Q_{\la+\epsilon_i}+\frac{1}{2}Q_{(\la,1)} \bigg),
    \label{eqn:DhL0}\\
( L_{1})^\perp\cdot Q_{\lambda}
=& \frac{1}{2}\sum_{i,j=1 \atop i\neq j}^{l(\la)} Q_{\la+\epsilon_i+\epsilon_j}
+\sum_{i=1}^{l(\la)}\frac{\la_i+2}{2} Q_{\la+2\epsilon_i}
+\frac{1}{2}\sum_{i=1}^{l(\la)} Q_{(\la+\epsilon_i,1)} \nonumber\\
&+\frac{1}{2} Q_{(\la,2)}
-\frac{1}{\hbar} \bigg(\sum_{i=1}^{l(\la)} Q_{\la+3\epsilon_i}
+\frac{1}{2}Q_{(\la,3)}-\frac{1}{2}Q_{(\la,2,1)}\bigg), \label{eqn:DhL1}\\
( L_{2})^\perp\cdot Q_{\lambda}
=&\sum_{i,j=1 \atop i\neq j}^{l(\la)} Q_{\la+3\epsilon_i+\epsilon_j}
+\sum_{i=1}^{l(\la)}\frac{\la_i+4}{2} Q_{\la+4\epsilon_i}
+\frac{1}{2}\sum_{i=1}^{l(\la)} Q_{(\la+3\epsilon_i,1)}
\nonumber\\
&+\frac{1}{2}\sum_{i=1}^{l(\la)} Q_{(\la+\epsilon_i,3)} -\frac{1}{2}\sum_{i=1}^{l(\la)} Q_{(\la+\epsilon_i,2,1)}
+Q_{(\la,4)}-\frac{1}{2}Q_{(\la,3,1)} \nonumber\\
&-\frac{1}{\hbar} \bigg(\sum_{i=1}^{l(\la)} Q_{\la+5\epsilon_i}
+\frac{1}{2}Q_{(\la,5)}-\frac{1}{2}Q_{(\la,4,1)} + \frac{1}{2}Q_{(\la,3,2)}\bigg). \label{eqn:DhL2}
\end{align}
}
\end{pro}

Now, we are ready to compute $\langle {L}_k^{(N)} \tau_{N}, \,\, Q_\mu \rangle$ for $k=0,1,2$. Essentially they are given  by the following functions of $\mu$:

\begin{align}
\Phi(\mu):=& \left( 8 |\mu| + \theta(1) \right)B_{\mu}
-\sum_{i=1}^{l(\mu)} \theta(\mu_i+1) B_{\mu+\epsilon_i}
-\frac{1}{2} \theta(1) B_{(\mu,1)}, \label{eqn:Phi}\\
\Psi(\mu):=&
 \sum_{i,j=1 \atop i\neq j}^{l(\mu)}  \theta(\mu_i+1)\theta(\mu_j+1)B_{\mu+\epsilon_i+\epsilon_j}
+ \theta_{(2)} B_{(\mu,2)} \nonumber\\
&+\sum_{i=1}^{l(\mu)} (\mu_i+2) \theta^{[2]}(\mu_i) B_{\mu+2\epsilon_i}
+\sum_{i=1}^{l(\mu)}  \theta(1)\theta(\mu_i+1)B_{(\mu+\epsilon_i,1)} \nonumber\\
&-\frac{1}{16}\bigg(2 \sum_{i=1}^{l(\mu)} \theta^{[3]}(\mu_i) B_{\mu+3\epsilon_i}
+ \theta_{(3)}B_{(\mu,3)}- \theta_{(2,1)}  B_{(\mu,2,1)}\bigg), \label{eqn:Psi}
\end{align}
and
\begin{align}\label{eqn:Ga}
\Gamma(\mu)
:=& 2 \sum_{i,j=1 \atop i\neq j}^{l(\mu)} \theta(\mu_j+1) \theta^{[3]}(\mu_i) B_{\mu+3\epsilon_i+\epsilon_j}
+  \sum_{i=1}^{l(\mu)} (\mu_i+4) \theta^{[4]}(\mu_i) B_{\mu+4\epsilon_i} \nonumber\\
&+  \sum_{i=1}^{l(\mu)}  \theta(1) \theta^{[3]}(\mu_i) \, B_{(\mu+3\epsilon_i,1)}
+ \sum_{i=1}^{l(\mu)}  \theta_{(3)} \theta(\mu_i+1) \, B_{(\mu+\epsilon_i,3)} \nonumber\\
&-  \sum_{i=1}^{l(\mu)}  \theta_{(2,1)} \theta(\mu_i+1) \, B_{(\mu+\epsilon_i,2,1)}
+ 2 \theta_{(4)} B_{(\mu,4)}
- \theta_{(3,1)}  B_{(\mu,3,1)} \nonumber\\
&-\frac{1}{16} \cdot\bigg(2 \sum_{i=1}^{l(\mu)}  \theta^{[5]}(\mu_i) B_{\mu+5\epsilon_i}
                            + \sum_{r=0}^2(-1)^r  \theta_{(5-r,r)} B_{(\mu,5-r,r)}\bigg),
\end{align}
where
\[ \theta^{[k]}(r):=\prod_{j=1}^k \theta(r+j)
\]
for all integers $k \geq 1$ and $r \geq 0$.
Note that $(k)$ is considered as a partition with only one part. So by definition, $\theta_{(k)} = \prod_{j=1}^k \theta(j)$,
 which is different from $\theta(k)$. For convenience, we also set $\theta_{(0)}=1$ and
 $\theta_{(\la, 0)}=\theta_{(0, \la)}=\theta_{\la}$
 for all positive partitions $\la$.

\begin{thm}\label{thm:hLtoPhi}
	For all strict partitions $\mu$,
	\begin{align*}
	\langle  L_{0}^{(N)}\cdot\tau_{N}, \,\, Q_{\mu} \rangle
	\,\, = \,\, & \left( \frac{\hbar}{16} \right)^{|\mu|} \, \frac{\theta_\mu}{16} \cdot\Phi(\mu),\\
	\langle  L_{1} \cdot\tau_{N}, \,\, Q_{\mu} \rangle
	\,\, = \,\, & \left( \frac{\hbar}{16} \right)^{|\mu|+2} \, \frac{\theta_\mu}{2} \cdot\Psi(\mu),\\
	\langle  L_{2} \cdot\tau_{N}, \,\, Q_{\mu} \rangle
	\,\, = \,\, & \left( \frac{\hbar}{16} \right)^{|\mu|+4} \, \frac{\theta_\mu}{2} \cdot\Gamma(\mu).
	\end{align*}
\end{thm}
{\bf Proof}:
The following formula was proved in \cite{LY} Lemma 4.1:  For any partition $\mu$,
\begin{equation} \label{eqn:inner-consis}
 \left\langle \sum_{\la \in DP} 2^{-l(\la)} f(\la) Q_{\la}, \,\, Q_{\mu} \right\rangle = f(\mu),
\end{equation}
where $f(\la)$ is any function of $\la$ such that
 $f(\la)$ is skew symmetric with respect to permutations of two parts of $\la$ which are not both $0$ and
$f((\la,0))=f(\la)$ for all partitions $\la$.
 Set \[ f(\la) := \left( \frac{\hbar}{16} \right)^{|\la|} \theta_\la B_\lambda.\]
 Then $f(\la)$ satisfies the above condition, and
 \[ \tau_N = \sum_{\la \in DP} 2^{-l(\la)} f(\la) Q_{\la}.\]
Hence by equation \eqref{eqn:inner-consis}, we have
\[ \langle \tau_{N}, \,\, Q_\mu \rangle \,\, = \,\, f(\mu) \]
for any partition $\mu$. In particular, we can use this formula to compute
\[\langle{L}_k^{(N)} \tau_{N}, \,\, Q_\mu \rangle \,\, = \,\, \langle \tau_{N}, \,\, ({L}_k^{(N)})^\perp Q_\mu\rangle \]
for $k=0, 1, 2$ and $\mu \in DP$. By Proposition \ref{pro:DhL},  we have
\begin{align}
\langle  L_{0}^{(N)}\cdot \tau_{N}, \,\, Q_{\mu} \rangle
=&  \left( \frac{|\mu|}{2}+\frac{1-4N^2}{16} \right) f(\mu)
  - \frac{1}{\hbar} \bigg(\sum_{i=1}^{l(\mu)} f(\mu+\epsilon_i) + \frac{1}{2} f((\mu,1)) \bigg),
                   \label{eqn:L0Inn} \\
\langle  L_{1} \cdot \tau_{N}, \,\, Q_{\mu} \rangle
=& \frac{1}{2} \bigg( \sum_{i,j=1 \atop i \neq j}^{l(\mu)} f(\mu+\epsilon_i+\epsilon_j)
            + \sum_{i=1}^{l(\mu)} (\mu_i+2) f(\mu+2\epsilon_i)   \nonumber \\
& \hspace{80pt}            + \sum_{i=1}^{l(\mu)} f((\mu+\epsilon_i,1))
            + f((\mu,2))   \bigg) \nonumber \\
&  - \frac{1}{\hbar} \bigg( \sum_{i=1}^{l(\mu)} f(\mu+3\epsilon_i)
   + \frac{1}{2} f((\mu,3))
   -\frac{1}{2} f((\mu,2,1))\bigg),
   \label{eqn:L1Inn}
\end{align}
and
\begin{align}
\langle  L_{2} \cdot \tau_{N}, \,\, Q_{\mu} \rangle
=& \sum_{i,j=1 \atop i\neq j}^{l(\mu)} f(\mu+3\epsilon_i+\epsilon_j)
+ \sum_{i=1}^{l(\mu)} \frac{\mu_i+4}{2} f(\mu+4\epsilon_i)
+ \frac{1}{2}\sum_{i=1}^{l(\mu)} f((\mu+3\epsilon_i,1))
\nonumber\\
&  +\frac{1}{2}\sum_{i=1}^{l(\mu)} f((\mu+\epsilon_i,3))
  - \frac{1}{2} \sum_{i=1}^{l(\mu)} f((\mu+\epsilon_i,2,1))
  +f((\mu,4))-\frac{1}{2} f((\mu,3,1))  \nonumber\\
& -\frac{1}{\hbar} \bigg( \sum_{i=1}^{l(\mu)} f(\mu+5\epsilon_i)
               + \frac{1}{2}  f((\mu,5)) - \frac{1}{2} f((\mu,4,1))+ \frac{1}{2} f((\mu,3,2)) \bigg).
    \label{eqn:L2Inn}
\end{align}
Note that $f(\mu)$ contains a factor $\theta_{\mu}$, and
\[ \theta_{(\mu, a_1, \cdots, a_k)}=\theta_{\mu} \, \theta_{(a_1, \cdots, a_k)},
\hspace{20pt} \theta_{\mu+k\epsilon_i} = \theta_\mu \, \theta^{[k]}(\mu_i)
\]
for any integers $k \geq 1$ and $a_1, \cdots, a_k \geq 0$.
After factoring out suitable factors from the right hand sides of equations \eqref{eqn:L0Inn}, \eqref{eqn:L1Inn}, and \eqref{eqn:L2Inn},
we obtain the desired formulas.
$\Box$

By Theorem \ref{thm:hLtoPhi}, to prove Virasoro constraints for $\tau_{N}$, we only need to show $\Phi(\mu)=0$, $\Psi(\mu)=0$, and $\Gamma(\mu)=0$ for all strict partitions $\mu$. These equations will be proved in Theorem \ref{thm:AhL0}, Theorem \ref{thm:AhL1} and Theorem \ref{thm:AhL2} respectively.

\subsection{$ L_{0}^{(N)}$ constraint}

In this subsection, we will prove the following theorem, which implies the $L_0^{(N)}$ constraint for $\tau_{N}$.
\begin{thm} \label{thm:AhL0}
	Let $\Phi(\mu)$ be the function defined by equation \eqref{eqn:Phi}. We have
    \[\Phi(\mu)=0 \]
    for all positive partitions $\mu$.
\end{thm}
{\bf Proof}:
We first simplify $\Phi(\mu)$ using the following formula:
\begin{align}\label{eqn:(la,1)}
B_{(\la,1)}=2B_{\la}-2\sum_{i=1}^{l(\la)} B_{\la+\epsilon_i}
\end{align}
for any partition $\la$. This formula is obtained
 by setting $\mathbf{t}=(1,0,0,\cdots)$ in equation \eqref{eqn:multi pr} with $r=1$.
Using this formula, we can remove $B_{(\mu,1)}$ in equation \eqref{eqn:Phi} and obtain
\begin{align} \label{eqn:Phi_sim}
\Phi(\mu)= 8 |\mu| B_{\mu} - 4 \sum_{i=1}^{l(\mu)} \mu_i (\mu_i+1) B_{\mu+\epsilon_i}.
\end{align}
Note that the right hand side of this equation does not depend on $N$ any more and all partitions involved have equal length.

Assume $\mu=(\mu_1,...,\mu_l)$. We prove this theorem by induction on $l$.

If $l=0$, then $\mu=\emptyset$ and $\Phi(\mu)=0$ holds trivially since $|\mu|=l(\mu)=0$.

If $l=1$,  $\Phi(\mu)$ is equal to
\begin{equation} \label{eqn:Phi_ini-1}
\Phi((\mu_1)) = 8 \mu_1 B_{(\mu_1)} - 4 \mu_1 (\mu_1+1) B_{(\mu_1 + 1)} = 0,
\end{equation}
where the last equality follows from equation \eqref{eqn:hook1}.

If $l=2$, $\Phi(\mu)$ is equal to
\begin{align}\label{eqn:Phi_ini}
\Phi((\mu_1,\mu_2))
&= 8(\mu_1+\mu_2) B_{(\mu_1,\mu_2)}
- 4 \mu_1 (\mu_1+1) B_{(\mu_1+1,\mu_2)}
- 4 \mu_2 (\mu_2+1) B_{(\mu_1,\mu_2+1)}  \nonumber \\
&=0,
\end{align}
where the last equality follows from straightforward calculations using formula \eqref{eqn:hook2}.

For any even integer $l>2$, we apply recursion formula \eqref{eqn:BRec2} to each term in $\Phi(\mu)$ to obtain
\begin{align} \label{eqn:phi_finish}
\Phi(\mu)=\sum_{j=2}^l (-1)^jB_{(\mu_1,\mu_j)}\Phi(\mu^{\{1,j\}})
+\sum_{i=2}^l (-1)^iB(\mu^{\{1,i\}})\cdot\Phi((\mu_1,\mu_i)).
\end{align}

For any odd integer $l>1$, we first replace each partition $\nu$ appeared in
the right hand side of equation \eqref{eqn:Phi_sim} by $(\nu, 0)$, then
apply equation \eqref{eqn:BRec2} to expand each term in $\Phi(\mu)$. The calculations
are similar to the $l$ even case except that an extra term
\[B_{(\mu_1)} \Phi (\mu^{\{1\}})+ B_{\mu^{\{1\}}} \Phi((\mu_1))  \]
should be added to the right hand side of equation
\eqref{eqn:phi_finish}.

Since the lengths of $\mu^{\{1\}}$ and $\mu^{\{1,i\}}$ are less than $l$, the theorem is reduced to the cases
of $l=1$ and $l=2$, which have been considered in equations \eqref{eqn:Phi_ini-1} and \eqref{eqn:Phi_ini}.
The theorem is thus proved.
$\Box$

\subsection{$L_{1}$ constraint}

In this subsection, we will prove $\Psi(\mu)=0$, which is equivalent to the $L_1$ constraint of $\tau_{N}$. We will need the following
two lemmas.
\begin{lem}\label{lem:N1}
For any partition $\mu=(\mu_1,\mu_2,...,\mu_l)$ with $l \geq 3$, define
\begin{align}\label{eqn:N1def}
M_1(\mu):=\sum_{i,j=2 \atop i\neq j}^l (-1)^iB_{(\mu+\epsilon_j)^{\{1,i\}}}\cdot \omega(\mu_1,\mu_i,\mu_j),
\end{align}
where
\begin{align} \label{eqn:omega1}
\omega(\mu_1,\mu_i,\mu_j):=&
                   a_1(\mu_1, \mu_j) B_{(\mu_1+1,\mu_i)}
                 + a_1(\mu_i, \mu_j) B_{(\mu_1,\mu_i+1)},
\end{align}
and
\begin{align} \label{eqn:akm}
a_1(k,m):=&\theta(k+1)\theta(m+1)-\theta(1)\big\{\theta(k+1)+\theta(m+1)\big\}
\end{align}
for all non-negative integers $k$ and $m$.
Then
\[ M_1(\mu)=0\]
 for all weakly positive partitions $\mu$ with even length.
\end{lem}
{\bf Proof:}
For $j \neq 1 , i$, we use recursion formula \eqref{eqn:BRec2j} to expand each term $B_{(\mu+\epsilon_j)^{\{1,i\}}}$
in the definition of $M_1(\mu)$ to obtain
\begin{equation} \label{eqn:expand_N1}
B_{(\mu+\epsilon_j)^{\{1,i\}}}=\sum_{m=2 \atop m \neq i, j}^l (-1)^{\tilde{j}(i)+\tilde{m}(i,j)}  B_{(\mu_j+1,\mu_m)} B_{\mu^{\{1,i,j,m\}}},
\end{equation}
where
\begin{equation} \label{eqn:tilde}
\tilde{j}(i):=j-\delta_{j<i}, {\rm \,\,\, and \,\,\, } \tilde{m}(i,j):=m-\delta_{m<i}-\delta_{m>j}.
\end{equation}

After the expansion, we can compute all factors of type $B_{\nu}$ with $l(\nu)=2$
using equation \eqref{eqn:hook2} and obtain
\begin{align} \label{eqn:M1Rec}
M_1(\mu)=&
\sum_{\{a,b,c\} \subseteq \{2,...,l\}\atop a<b<c} \frac{2^{\mu_1+\mu_a+\mu_b+\mu_c+2} B_{\mu^{\{1,a,b,c\}}}}{(\mu_1+1)!(\mu_a+1)!(\mu_b+1)!(\mu_c+1)!} \sum_{(i,j,m)\in P(a,b,c)} (-1)^{i+\tilde{j}(i)+\tilde{m}(i,j)}
            \nonumber \\
&\hspace{60pt} \cdot \big\{ 16\mu_1^2\cdot\rho_2(\mu_i,\mu_j,\mu_m)
                            +\mu_1\cdot\rho_1(\mu_i,\mu_j,\mu_m)
                            +\rho_0(\mu_i,\mu_j,\mu_m)\big\},
\end{align}
where $P(a,b,c)$ denotes the set of all permutations of $(a,b,c)$, and
\begin{align} \label{eqn:defrho}
\rho_0(\mu_i,\mu_j,\mu_m)=&
            C_{j,m} \,\, \mu_i \left\{-16\mu_j(\mu_j+1)(\mu_i+1)+\theta(1)^2\right\},
            \nonumber \\
\rho_1(\mu_i,\mu_j,\mu_m)=&C_{j,m} \, \left\{ 16\mu_j(\mu_j+1)(1-\mu_i^2)-\theta(1)^2 \right\},
            \nonumber \\
\rho_2(\mu_i,\mu_j,\mu_m)=&C_{j,m} \, (\mu_i+1)\mu_j(\mu_j+1),
\end{align}
with \[ C_{j,m}:=\frac{(\mu_m+1)(\mu_j+1-\mu_m)}{\mu_j + \mu_m + 1 }. \hspace{20pt} \]

 In Lemma \ref{lem:rho} in the appendix, we will prove three elementary identities
 \[ \sum\limits_{(i,j,m)\in P(a,b,c)} (-1)^{i+\tilde{j}(i)+\tilde{m}(j,i)}\rho_k(\mu_i,\mu_j,\mu_m)=0 \]
 for $k=0, 1, 2$. This lemma follows from these identities.
$\Box$

\begin{lem}\label{lem:N2}
	For any partition $\mu=(\mu_1,\mu_2,...,\mu_l)$ with $l\geq 3$, define
	\begin{align}\label{eqn:N2def}
	M_2(\mu):= M_1(\mu) + \sum_{i=2}^l B_{(\mu+\epsilon_i)^{\{1\}}}\cdot \tilde\omega(\mu_1,\mu_i),
	\end{align}
	where
	\begin{align} \label{eqn:omega1tilde}
        \tilde\omega(\mu_1,\mu_i):=&  a_1(\mu_1, \mu_i) B_{(\mu_1+1)},
	\end{align}
    with $a_1(\mu_1, \mu_i)$  defined by equation \eqref{eqn:akm}, and $M_1(\mu)$ is defined by equation \eqref{eqn:N1def}.
	Then \[ M_2(\mu)=0 \]
 for all positive partitions $\mu$ with odd length.
\end{lem}
{\bf Proof:}
As in the proof of Lemma \ref{lem:N1}, We first use recursion formula \eqref{eqn:BRec2j} to expand $M_1(\mu)$
with respect to the part $(\mu_j+1)$.
Since all partitions $\nu$ involved  have odd length, we need replace $\nu$ by $(\nu, 0)$ before doing expansion.
This will produce extra terms for the expansion of $M_1(\mu)$. More precisely,
we need add an extra term
\[(-1)^{\tilde{j}(i)+1}B_{(\mu_j+1)}B_{\mu^{\{1,i,j\}}}\]
 to the right hand side of equation \eqref{eqn:expand_N1}, and an extra term
 \[ \sum_{i,j=2 \atop i\neq j}^l (-1)^{i+\tilde{j}(i)+1} B_{(\mu_j+1)} B_{\mu^{\{1,i,j\}}} \omega(\mu_1,\mu_i,\mu_j) \]
 should be added to the right hand side of equation \eqref{eqn:M1Rec}.

On the other hand, we also use recursion formula \eqref{eqn:BRec2j} to expand the second part in the definition of $M_2(\mu)$ with respect to the part $(\mu_i+1)$ and obtain
\[\sum_{i=2}^l B_{(\mu+\epsilon_i)^{\{1\}}}\cdot \tilde\omega(\mu_1,\mu_i)
=\sum_{i,j=2 \atop i\neq j}^l (-1)^{i+\tilde{j}(i)+1} B_{(\mu_i+1,\mu_j)} B_{\mu^{\{1,i,j\}}} \cdot \tilde\omega(\mu_1,\mu_i).\]

As in the proof of Lemma \ref{lem:N1}, we can still use Lemma \ref{lem:rho} to show that
the contribution from the right hand side of equation \eqref{eqn:M1Rec} is $0$. The remaining terms in $M_2(\mu)$ are
\begin{equation} \label{eqn:M2Rec}
M_2(\mu)=\sum_{i,j=2 \atop i\neq j}^l (-1)^{i+\tilde{j}(i)+1}B_{\mu^{\{1,i,j\}}} \Big(B_{(\mu_j+1)}\omega(\mu_1,\mu_i,\mu_j)+B_{(\mu_i+1,\mu_j)}\tilde\omega(\mu_1,\mu_i) \Big).
\end{equation}
Note that
\begin{equation} \label{eqn:akm2}
a_1(k,m)=16km(k+1)(m+1)-\theta(1)^2,
\end{equation}
which is included in the definition of $\omega$ and $\tilde\omega$.
 After separating terms containing the factor $\theta(1)^2$ from terms  not containing this factor, a straightforward calculation using
 above equation and equations \eqref{eqn:hook1} and  \eqref{eqn:hook2}
 shows
\begin{align*}
& B_{(\mu_j+1)}\omega(\mu_1,\mu_i,\mu_j)+B_{(\mu_i+1,\mu_j)} \tilde\omega(\mu_1,\mu_i)   \\
=& \frac{2^{\mu_1+\mu_i+\mu_j+6}}{\mu_1 ! \mu_i ! \mu_j !} \,\,
            \bigg(\frac{\mu_1 (\mu_i^2+\mu_i+\mu_j^2+\mu_j)}{\mu_i+\mu_j+1} - \mu_i \mu_j \bigg)  \\
& -\frac{\theta(1)^2 \, 2^{\mu_1+\mu_i+\mu_j+2}}{(\mu_1+1)!(\mu_i+1)!(\mu_j+1)!} \,\,
            \bigg( \mu_1 + \frac{1-\mu_i^2-\mu_j^2}{1+\mu_i+\mu_j} \bigg).
\end{align*}
Note that this expression is symmetric with respect to $i$ and $j$. Since $(-1)^{i+\tilde{j}(i)}$ is skew symmetric with respect to $i$ and $j$, equation \eqref{eqn:M2Rec} implies that $M_2(\mu)=0$.
$\Box$

We are now ready to prove
\begin{thm} \label{thm:AhL1}
Let $\Psi(\mu)$ be the function of $\mu$ defined by equation \eqref{eqn:Psi}.
We have \[\Psi(\mu)=0 \]
for all strict partitions $\mu$.
In particular, $\tau_{N}$ satisfies the $L_1$ constraint.
\end{thm}
{\bf Proof}:
We first remove terms of form $B_{(\la, 1)}$ in $\Psi(\mu)$ using equation \eqref{eqn:(la,1)} and remove terms of form
$B_{(\la, 2,1)}$ using  following formula
\begin{align}\label{eqn:(la,2,1)}
B_{(\la,2,1)}=2\sum_{i=1}^{l(\la)} B_{\la+3\epsilon_i}+ B_{(\la,3)},
\end{align}
which is obtained by evaluating both sides of equation \eqref{eqn:multi pr} with $r=3$ at the point $\mathbf{t}=(1,0,0, \cdots)$.
After simplification, $\Psi(\mu)$ can be written as
\begin{align}\label{eqn:Psi_simp}
\Psi(\mu):=&
\sum_{i,j=1 \atop i\neq j}^{l(\mu)} a_1 (\mu_i, \mu_j) B_{\mu+\epsilon_i+\epsilon_j}
+\theta_{(2)} B_{(\mu,2)} +2 \sum_{i=1}^{l(\mu)} \theta(1) \theta(\mu_i+1) B_{\mu+\epsilon_i} \nonumber\\
&+\sum_{i=1}^{l(\mu)} a_2(\mu_i) B_{\mu+2\epsilon_i}
 -\frac{1}{8} \sum_{i=1}^{l(\mu)} a_3(\mu_i) B_{\mu+3\epsilon_i}
                    -  \frac{3}{2}\theta_{(2)} B_{(\mu,3)},
\end{align}
where $a_1(k,m)$ is defined by equation \eqref{eqn:akm} and
\begin{eqnarray} \label{eqn:a3k}
a_2(k) &:=& (k+2)\theta^{[2]}(k)-2\theta(1) \theta(k+1), \nonumber \\
a_3(k) &:=&  \theta^{[3]}(k)-\theta_{(2,1)}
\end{eqnarray}
for any non-negative integer $k$.

Assume $\mu=(\mu_1,...,\mu_l)$.
We prove this theorem by induction on $l$.

{\bf Step 1}: Prove $\Psi(\mu)=0$ if $l=0, 1, 2$.

If $l=0$, we have
\begin{align}\label{eqn:Psi-emp}
\Psi(\emptyset)
= \theta_{(2)} \Big(B_{(2)} - \frac{3}{2} \, B_{(3)} \Big)=0,
\end{align}
since $B_{(2)}=2$ and $B_{(3)}=\frac{4}{3}$.

If $l=1$, we have
\begin{align}\label{eqn:Psil=1-ini}
\Psi((\mu_1))
=& \theta_{(2)} B_{(\mu_1,2)} + 2 \theta(1)\theta(\mu_1+1)B_{(\mu_1+1)} + a_2 (\mu_1) B_{(\mu_1+2)}
        \nonumber\\
& -\frac{1}{8} a_3(\mu_1)  B_{(\mu_1+3)} -\frac{3}{2}  \theta_{(2)}B_{(\mu_1,3)} \nonumber\\
=&0,
\end{align}
where the last equality follows from straightforward calculations using equations \eqref{eqn:hook1} and \eqref{eqn:hook2}.

If $l=2$, we first remove $B_{(\mu_1,\mu_2,2)}$ and $B_{(\mu_1,\mu_2,3)}$ in $\Psi((\mu_1, \mu_2))$ using formula
\begin{align*}
B_{(\mu_1,\mu_2,k)}=& B_{(\mu_1)}B_{(\mu_2,k)} - B_{(\mu_2)} B_{(\mu_1,k)} + B_{(k)} B_{(\mu_1,\mu_2)}
\end{align*}
for any positive integer $k$, which is obtained using equation \eqref{eqn:BRecOdd}.
We then have
\begin{align}\label{eqn:Psil2_expan}
\Psi((\mu_1,\mu_2))=g_1(\mu_1,\mu_2)+g_2(\mu_1,\mu_2) + \Psi(\emptyset) B_{(\mu_1, \mu_2)},
\end{align}
where
\begin{align*}
g_1(\mu_1,\mu_2)
:=& 2 a_1(\mu_1, \mu_2) B_{(\mu_1+1,\mu_2+1)}
+ \theta_{(2)} \left\{ B_{(\mu_1)}B_{(\mu_2,2)} - B_{(\mu_2)} B_{(\mu_1,2)} \right\} \\
& +2 \theta(1)\theta(\mu_1+1) B_{(\mu_1+1,\mu_2)} +2 \theta(1)\theta(\mu_2+1) B_{(\mu_1,\mu_2+1)} \nonumber\\
& + a_2(\mu_1) B_{(\mu_1+2,\mu_2)} + a_2(\mu_2) B_{(\mu_1,\mu_2+2)},
\end{align*}
and
\begin{align*}
g_2(\mu_1,\mu_2)
:=&-  \frac{3}{2} \theta_{(2)} \left\{ B_{(\mu_1)}B_{(\mu_2,3)} - B_{(\mu_2)} B_{(\mu_1,3)} \right\} \\
&   - \frac{1}{8} \left\{  a_3(\mu_1) B_{(\mu_1+3,\mu_2)} +a_3(\mu_2) B_{(\mu_1,\mu_2+3)} \right\}.
\end{align*}

By straightforward calculations using equations \eqref{eqn:hook1} and \eqref{eqn:hook2}, we have
\begin{align*}
& g_1(\mu_1,\mu_2)=-g_2(\mu_1,\mu_2)\\
=& \frac{2^{\mu_1+\mu_2+4}(\mu_1-\mu_2)}{\mu_!!\mu_2!}\cdot \{-12N^2+4(\mu_1^2+\mu_2^2)+12(\mu_1+\mu_2)-4\mu_1\mu_2+11\}.
\end{align*}
Hence
\begin{align} \label{eqn:Psi-ini}
\Psi((\mu_1,\mu_2))=0.
\end{align}

{\bf Step 2}: Prove $\Psi(\mu)=0$ if $l$ is an even integer bigger than $2$.

We use  recursion formula \eqref{eqn:BRec2} to expand $\Psi(\mu)$ in equation \eqref{eqn:Psi_simp}. For those partitions $\nu$ with odd length, we need to replace them by $(\nu,0)$ before applying the recursion formula \eqref{eqn:BRec2}. If after the first expansion, we obtain some terms containing a factor
\begin{align}\label{eqn:Psi_excep}
B_{(\mu,s)^{\{1\}}}, \text{\ for\ }s\in\{0,2,3\},
\end{align}
we will use formula \eqref{eqn:BRec2j} with respect to the $l$-th part to expand them again.
After such expansions, we obtain
\begin{align}\label{eqn:Psitow12}
\Psi(\mu)
=&\sum_{i=2}^l (-1)^iB_{(\mu_1,\mu_i)}\Psi(\mu^{\{1,i\}})
   +M_1(\mu) \nonumber \\
& +\sum_{i=2}^l (-1)^iB_{\mu^{\{1,i\}}}\cdot \{ \Psi((\mu_1,\mu_i))-B_{(\mu_1,\mu_i)}\cdot\Psi(\emptyset) \},
\end{align}
where $M_1(\mu)$ is defined by equation \eqref{eqn:N1def}.
Since $M_1(\mu)=0$ by Lemma \ref{lem:N1}, this reduces the proof of $\Psi(\mu)=0$ to the $l=0$ and $l=2$ cases, which have
been  considered
in equations \eqref{eqn:Psi-emp} and  \eqref{eqn:Psi-ini} respectively.

{\bf Step 3}: Prove $\Psi(\mu)=0$ if $l$ is an odd integer bigger than $1$.

 As in step 2, we expand  $\Psi(\mu)$  by recursion formula \eqref{eqn:BRec2}. For those partitions $\nu$ appeared in the right hand side of equation \eqref{eqn:Psi_simp} which have odd length, we need to replace them by $(\nu,0)$ before applying recursion formula \eqref{eqn:BRec2}. This will produce some extra terms containing the factor $B_{(\mu_1, 0)}$.
  Since $(\mu, 2)$ and $(\mu, 3)$ have even length, expansion of corresponding terms do not produce
  such factors. After factoring out $B_{(\mu_1, 0)}$ from these terms, we obtain an expression
  which coincides with most terms of $\Psi(\mu^{\{1\}})$ but with terms
  $\theta_{(2)} B_{(\mu^{\{1\}},2)} - \frac{3}{2}\theta_{(2)} B_{(\mu^{\{1\}},3)}$ missing.
  We can express summation of such terms as
 \[ B_{(\mu_1,0)} \left\{ \Psi(\mu^{\{1\}}) - \theta_{(2)} B_{(\mu^{\{1\}},2)} +
            \frac{3}{2}\theta_{(2)} B_{(\mu^{\{1\}},3)} \right\}. \]
 We then expand $B_{(\mu^{\{1\}},k)}= B_{(\mu^{\{1\}},k,0)}$
 using recursion formula \eqref{eqn:BRec2j} with $j=l$ for $k=2, 3$.
After regrouping terms, we obtain
\begin{align}\label{eqn:Psitotw12}
\Psi(\mu)
=&\sum_{i=2}^l (-1)^iB_{(\mu_1,\mu_i)}\Psi(\mu^{\{1,i\}})
+B_{(\mu_1,0)}\Psi(\mu^{\{1\}}) + B_{\mu^{\{1\}}}\cdot \left\{ \Psi((\mu_1)) - B_{(\mu_1)} \Psi(\emptyset) \right\} \nonumber\\
&+M_2(\mu)
+\sum_{i=2}^l (-1)^i B_{\mu^{\{1,i\}}}\cdot \left \{\Psi((\mu_1, \mu_i)) - B_{(\mu_1, \mu_i)} \Psi(\emptyset) \right\} \nonumber \\
& +  \theta_{(2)} \left( B_{(\mu_1, 2)} - \frac{3}{2} B_{(\mu_1, 3)} \right) \sum_{i=2}^l (-1)^i B_{\mu^{\{1,i\}}}  B_{(\mu_i)},
\end{align}
where $M_2(\mu)$ is defined by equation \eqref{eqn:N2def}, which is equal to $0$ by Lemma \ref{lem:N2}.
The last term on the right hand side of the above equation is also equal to $0$ by
Lemma \ref{lem:even-1} applied to $\la= \mu^{\{1\}}$.
By induction, the theorem is reduced to the cases discussed in step 1.
This theorem is thus proved.  $\Box$

\subsection{$L_{2}$ constraint}

In this subsection, we will prove $\Gamma(\mu)=0$, which is equivalent to the $L_2$ constraint for $\tau_{N}$.
We will need the following
\begin{lem}\label{lem:N}
	For any partition $\mu=(\mu_1,...,\mu_l)$ with $l\geq 3$, define
	\begin{align} \label{eqn:RDef}
	R(\mu) := & \sum_{i=2}^l (-1)^i (\mu_1+\mu_i)
        \Big(\sum_{j=2 \atop j\neq i}^{l} a_3(\mu_j) B_{(\mu+3\epsilon_j)^{\{1,i\}}}
	            +12\theta_{(2)}B_{(\mu,3)^{\{1,i\}}}\Big) B_{(\mu_1,\mu_i)} \nonumber \\
	&+\sum_{i=2}^l (-1)^i B_{\mu^{\{1,i\}}} \left\{ (|\mu|-\mu_1-\mu_i) \xi(\mu_1,\mu_i)
                - 12 (\mu_1+\mu_i)\theta_{(2)} B_{(3)} B_{(\mu_1,\mu_i)} \right\},
	\end{align}
	where $a_3(k)$ is defined by equation \eqref{eqn:a3k}, and
	\begin{align*}
	\xi(\mu_1,\mu_i):=a_3(\mu_1) B_{(\mu_1+3,\mu_i)}
	                 + a_3(\mu_i) B_{(\mu_1,\mu_i+3)}
	              + 12 \theta_{(2)} \left\{B_{(\mu_1)}B_{(\mu_i,3)}-B_{(\mu_1,3)}B_{(\mu_i)} \right\}.
	\end{align*}
	Then
	\[R(\mu)=0\]
    for all weakly positive partition $\mu$ with even length $l \geq 4$ such that $\mu_i>0$ for all $2 \leq i \leq l$.
\end{lem}
{\bf Proof:}
We first write the factor $|\mu|-\mu_1-\mu_i$ in the definition of $R(\mu)$ as
\[ |\mu|-\mu_1-\mu_i= \sum_{j=2 \atop j \neq  i}^l \mu_j.\]
We then use recursion formula \eqref{eqn:BRec2j} to expand following terms in $R(\mu)$ with respect to
parts which have different forms from other parts
\begin{align*}
B_{(\mu+3\epsilon_j)^{\{1,i\}}}=&\sum_{m=2 \atop m\neq i,m\neq j}^l
          (-1)^{\tilde{j}(i)+\tilde{m}(i,j)}B_{(\mu_j+3,\mu_m)}B_{\mu^{\{1,i,j,m\}}},\\
\sum_{j=2 \atop j\neq i}^l \mu_j B_{\mu^{\{1,i\}}}=&\sum_{j,m=2 \atop j\neq i,m\neq i,m\neq j}^l
                (-1)^{\tilde{j}(i)+\tilde{m}(i,j)}\mu_jB_{(\mu_j,\mu_m)}B_{\mu^{\{1,i,j,m\}}},\\
B_{(\mu,3,0)^{\{1,i\}}}=&\sum_{j,m=2 \atop j\neq i,m\neq i,m\neq j}^l(-1)^{\tilde{j}(i)+\tilde{m}(i,j)+1}
        B_{(\mu_j,3)}B_{(\mu_m,0)}B_{\mu^{\{1,i,j,m\}}}+B_{(3,0)}B_{\mu^{\{1,i\}}}.
\end{align*}
Note that partition $(\mu,3,0)^{\{1,i\}}$ is  weakly positive since the only possible zero part of $\mu$ is $\mu_1$ which is excluded
in this partition. When expanding $B_{(\mu,3,0)^{\{1,i\}}}$, we first expand them with respect to the part $3$, then expand again with respect to the zero part. The terms $B_{(3,0)}B_{\mu^{\{1,i\}}}$ in the last equation are cancelled with corresponding terms
in the second line of the definition of $R(\mu)$.

After above expansions,
we obtain
\begin{align} \label{eqn:RRec}
R(\mu)=&\sum_{i,j,m=2 \atop i\neq j,i\neq m,j\neq m}^l (-1)^{i+\tilde{j}(i)+\tilde{m}(i,j)} B_{\mu^{\{1,i,j,m\}}}
\cdot\Big\{\eta_1(\mu_1,\mu_i,\mu_j,\mu_m)+\eta_2(\mu_1,\mu_i,\mu_j,\mu_m)\Big\},
\end{align}
where
\begin{align*}
\eta_1(\mu_1,\mu_i,\mu_j,\mu_m) =& (\mu_1+\mu_i) B_{(\mu_1,\mu_i)}\Big(a_3(\mu_j)
             B_{(\mu_j+3,\mu_m)}-12\theta_{(2)}B_{(\mu_j,3)}B_{(\mu_m)}\Big),\\
\eta_2(\mu_1,\mu_i,\mu_j,\mu_m) =&\mu_jB_{(\mu_j,\mu_m)}\xi(\mu_1,\mu_i).
\end{align*}

By straightforward calculations using equations \eqref{eqn:hook1} and \eqref{eqn:hook2}, we have
\begin{align*}
\eta_1(\mu_1,\mu_i,\mu_j,\mu_m) &-\eta_1(\mu_1,\mu_i,\mu_m,\mu_j)
=\frac{2^{\mu_1+\mu_i+\mu_j+\mu_m+7}(\mu_1-\mu_i)(\mu_j-\mu_m)}{\mu_1!\mu_i!\mu_j!\mu_m!} \\
  &              \cdot  (-12N^2+4\mu_j^2+4\mu_m^2+12\mu_j+12\mu_m-4\mu_j\mu_m+11),\\
\eta_2(\mu_1,\mu_i,\mu_j,\mu_m) &-\eta_2(\mu_1,\mu_i,\mu_m,\mu_j)
=\frac{2^{\mu_1+\mu_i+\mu_j+\mu_m+7}(\mu_1-\mu_i)(\mu_j-\mu_m)}{\mu_1!\mu_i!\mu_j!\mu_m!} \\
&        \cdot  (-12N^2+4\mu_1^2+4\mu_i^2+12\mu_1+12\mu_i-4\mu_1\mu_i+11).
\end{align*}

Since $(-1)^{i+\tilde{j}(i)+\tilde{m}(i,j)} B_{\mu^{\{1,i,j,m\}}} $ is skew symmetric with respect to $j$ and $m$,
equation \eqref{eqn:RRec} then implies
\begin{align*}
R(\mu)=&\sum_{\{a,b,c\}\in\{2,...,l\}\atop a<b<c} \frac{2^{\mu_1+\mu_a+\mu_b+\mu_c+7}(-1)^{a+b+c+1}B_{\mu^{\{1,a,b,c\}}}}{\mu_1!\mu_a!\mu_b!\mu_c!}\\
&\ \ \ \ \ \ \ \ \ \ \cdot \sum_{(i,j,m)\in\sigma(a,b,c)}(\mu_1-\mu_i)(\mu_j-\mu_m) \big\{ h_{a,b,c}(\mu)-4(\mu_1\mu_i+\mu_j\mu_m)\big\},
\end{align*}
where $\sigma(a,b,c) := \{(a,b,c),(b,c,a),(c,a,b)\}$, and
\begin{align*}
h_{a,b,c}(\mu)=-24N^2+22+4(\mu_1^2+\mu_a^2+\mu_b^2+\mu_c^2)+12(\mu_1+\mu_a+\mu_b+\mu_c).
\end{align*}
The lemma then follows from elementary identities in  Lemma \ref{lem:3identities} in the appendix.
$\Box$

We are now ready to prove the following
\begin{thm} \label{thm:AhL2}
Let $\Gamma(\mu)$ be the function of $\mu$ defined by equation \eqref{eqn:Ga}. We have
\[\Gamma(\mu)=0\]
for all strict partitions $\mu$.
In particular, $\tau_{N}$ satisfies the $L_2$ constraint.
\end{thm}
{\bf Proof}:
We first use Theorem \ref{thm:AhL0} to simplify  $\Gamma(\mu)$ defined in equation \eqref{eqn:Ga}.

By Theorem \ref{thm:AhL0}, $\Phi((\mu,3))=0$ for any positive partition $\mu$. Hence we have
\[\sum_{i=1}^l \theta(\mu_i+1)B_{(\mu+\epsilon_i,3)}
=(8|\mu|+24+\theta(1))B_{(\mu,3)}-\theta(4)B_{(\mu,4)}-\frac{1}{2}\theta(1)B_{(\mu,3,1)}.\]
Since $\Phi((\mu,2,1))=0$, we have
\[\sum_{i=1}^l \theta(\mu_i+1)B_{(\mu+\epsilon_i,2,1)}
=(8|\mu|+24+\theta(1))B_{(\mu,2,1)}-\theta(3)B_{(\mu,3,1)}.\]
Since $\Phi(\mu+3\epsilon_i)=0$ for all $1 \leq i \leq l(\mu)$, we have
\begin{align*}
\sum_{j\neq i}^l \theta(\mu_j+1)B_{(\mu+3\epsilon_i+\epsilon_j)}
=(8|\mu|+24+\theta(1))B_{\mu+3\epsilon_i}
-\theta(\mu_i+4)B_{(\mu+4\epsilon_i)}
-\frac{1}{2}\theta(1)B_{(\mu+3\epsilon_i,1)}.
\end{align*}

We first use the above formulas to reduce corresponding summations in $\Gamma(\mu)$, then remove
$B_{(\mu, 2,1)}$ using equation \eqref{eqn:(la,2,1)}. We obtain the following simplification
of $\Gamma(\mu)$:
\begin{align}\label{eqn:Ga_simp}
\Gamma(\mu) =&\{8|\mu|+24+\theta(1)\}\Big(2\sum_{i=1}^{l(\mu)} a_3(\mu_i) B_{\mu+3\epsilon_i}
+24\theta_{(2)}B_{(\mu,3)}\Big) \nonumber\\
&+\sum_{i=1}^{l(\mu)}(\mu_i+2)\theta^{[4]}(\mu_i)B_{\mu+4\epsilon_i}
+\theta_{(4)}B_{(\mu,4)}
-\frac{1}{2}\theta_{(3,1)}B_{(\mu,3,1)} \nonumber\\
&-\frac{1}{16}\cdot\bigg(2\sum_{i=1}^{l(\mu)} \theta^{[5]}(\mu_i) B_{(\mu+5\epsilon_i)}
+\sum_{r=0}^2(-1)^r \theta_{(5-r,r)}B_{(\mu,5-r,r)}\bigg),
\end{align}
where $a_3(\mu_i)$ is defined by equation \eqref{eqn:a3k}.

Using the simple fact that
\[ \theta^{[3]}(0)-\theta_{(2,1)} = 24 \theta_{(2)}  {\rm \,\,\,\,\, and \,\,\,\,\, } \theta^{[k]}(0)=\theta_{(k)} \]
for all $k>0$, it is straightforward to show that the right hand side of equation  \eqref{eqn:Ga_simp} does not change value if
we replace $\mu$ by $(\mu, 0)$. Hence we have
$\Gamma(\mu)=\Gamma((\mu,0))$
for any partition $\mu$.
Since the right hand side of equation  \eqref{eqn:Ga_simp} is skew symmetric
with respect to permutations of weakly positive partitions, we have
$\Gamma(\mu)= \pm \Gamma((0,\mu))$
for any positive partition $\mu$.
Hence $\Gamma(\mu)=0$ if and only if $\Gamma((0,\mu))=0$.
In particular, if $\mu$ is a strict partition with odd length, instead of considering $\Gamma(\mu)$,
we will consider $\Gamma((0,\mu))$. In a summary,
 to prove $\Gamma(\mu)=0$ for all strict partitions $\mu$, it suffices to show that
$\Gamma(\mu)=0$ for all weakly positive partitions $\mu=(\mu_1, \mu_2, \cdots \mu_l)$
with $l$ even and $\mu_i>0$ for $2 \leq i \leq l$. We will prove this fact by induction on $l$.

If $l=0$, we have
\begin{align}\label{eqn:Ga_l=0ex}
\Gamma(\emptyset)=c_1+c_2,
\end{align}
where
\begin{align*}
c_1:=&24\{24+\theta(1)\}\theta_{(2)}B_{(3)}
+\theta_{(4)}B_{(4)}
-\frac{1}{2}\theta_{(3,1)}B_{(3,1)},\\
c_2:=&-\frac{1}{16}\sum_{r=0}^2(-1)^r \theta_{(5-r,r)}B_{(,5-r,r)}.
\end{align*}
A straightforward calculation using equations \eqref{eqn:hook1} and \eqref{eqn:hook2} shows that
$c_1=-c_2 = 64 \theta_{(3)}.$
Hence
\begin{align}\label{eqn:Ga_l=0}
\Gamma(\emptyset)=0.
\end{align}

If $l=2$, we use equations \eqref{eqn:BRecOdd} and \eqref{eqn:BRec2} to expand
$B_\nu$ occurred in $\Gamma((\mu_1, \mu_2))$ with $l(\nu)=3$ or $4$   and
obtain
\begin{align}\label{eqn:Ga_l=2ex}
\Gamma((\mu_1,\mu_2))=f_1(\mu_1,\mu_2) + f_2(\mu_1,\mu_2)+B_{(\mu_1, \mu_2)} \Gamma(\emptyset),
\end{align}
where
\begin{align*}
f_1(\mu_1,\mu_2):=&  \{8(\mu_1+\mu_2)+24+\theta(1)\}\Big( 2 a_3(\mu_1) B_{(\mu_1+3,\mu_2)}
                    +  2 a_3(\mu_2) B_{(\mu_1,\mu_2+3)} \nonumber\\
& \hspace{20pt}+ 24 \theta_{(2)} \left( B_{(\mu_1)} B_{(\mu_2,3)} - B_{(\mu_2)} B_{(\mu_1,3)}  \right) \Big)
     + 192 (\mu_1+\mu_2) \theta_{(2)} B_{(3)} B_{(\mu_1, \mu_2)} \nonumber\\
&+(\mu_1+2)\theta^{[4]}(\mu_1)B_{(\mu_1+4,\mu_2)}
+(\mu_2+2)\theta^{[4]}(\mu_2)B_{(\mu_1,\mu_2+4)} \nonumber\\
&+\theta_{(4)} \left( B_{(\mu_1)} B_{(\mu_2,4)} - B_{(\mu_2)} B_{(\mu_1,4)} \right)
-\frac{1}{2}\theta_{(3,1)} \left( B_{(\mu_1, 1)} B_{(\mu_2,3)} - B_{(\mu_2, 1)} B_{(\mu_1,3)}  \right), \nonumber\\
f_2(\mu_1,\mu_2):=&-\frac{1}{8} \bigg(\theta^{[5]}(\mu_1) B_{(\mu_1+5,\mu_2)} + \theta^{[5]}(\mu_2) B_{(\mu_1,\mu_2+5)} \bigg)
                 \nonumber\\
& -\frac{1}{16} \sum_{r=0}^2(-1)^r \theta_{(5-r,r)} \left( B_{(\mu_1, r)} B_{(\mu_2,5-r)} - B_{(\mu_2, r)} B_{(\mu_1,5-r)} \right).
\end{align*}
A straightforward calculation using equations \eqref{eqn:hook1} and \eqref{eqn:hook2} shows that
\begin{align*}
&f_1(\mu_1,\mu_2)=-f_2(\mu_1,\mu_2) \\
=&\frac{2^{\mu_1+\mu_2+9}(\mu_1-\mu_2)}{\mu_1!\mu_2!}\sum_{k=1}^2
        \Big\{40N^4-20(2\mu_k^2+10\mu_k-\mu_1\mu_2+11)N^2\\
&  \hspace{60pt}             + 2\big(4\mu_k^4+40\mu_k^3+145\mu_k^2+225\mu_k\big)- \mu_1\mu_2(8\mu_k^2-4\mu_1\mu_2-55)+\frac{297}{2}\Big\}.
\end{align*}
Hence
\begin{align}\label{eqn:Ga_l=2}
\Gamma((\mu_1,\mu_2))=0.
\end{align}

For any even integer $l>2$, we use recursion formula \eqref{eqn:BRec2} to expand $\Gamma(\mu)$ in equation~\eqref{eqn:Ga_simp}.
For those partitions $\nu$ occurred in $\Gamma(\mu)$ with odd length, we need to replace it by $(\nu,0)$ before applying recursion formula \eqref{eqn:BRec2}.
Since $\nu$ is weakly positive, Lemma \ref{lem:odd0} guarantees equation \eqref{eqn:BRec2} can still be used for $(\nu,0)$
in this case.
After the first expansion, we obtain some terms containing a factor $B_{(\mu,s)^{\{1\}}}$ with $0 \leq s \leq 5$.
We can use formula~\eqref{eqn:BRec2j} to expand such terms again with respect to the $l$-th part.
After such expansions, we obtain
\begin{align*} 
\Gamma(\mu)
=&\sum_{i=2}^l (-1)^i B_{(\mu_1,\mu_i)}\Gamma(\mu^{\{1,i\}})
+\sum_{i=2}^l (-1)^i B_{\mu^{\{1,i\}}} \left\{ \Gamma((\mu_1,\mu_i))-B_{(\mu_1,\mu_i)}\Gamma(\emptyset) \right\}
+16R(\mu),
\end{align*}
where $R(\mu)$  is defined by equation \eqref{eqn:RDef} and it is equal to $0$ by  Lemma \ref{lem:N}.
By induction, the theorem is reduced to the $l=0$ and $l=2$ cases, which have been
considered in equations \eqref{eqn:Ga_l=0} and \eqref{eqn:Ga_l=2} respectively.
The theorem is thus proved.
$\Box$

{\bf Proof of Theorems \ref{thm:A=BGW} and \ref{thm:A=gBGW}}:
Since $L_0^{(N)}$, $L_1$, $L_2$ generate all Virasoro operators $L_m^{(N)}$ for $m \geq 0$,
Theorems \ref{thm:AhL0}, \ref{thm:AhL1}, \ref{thm:AhL2} imply that $\tau_N$ satisfies the Virasoro
constraints~\eqref{eqn:VirConst} for all $N$. Moreover,
since $Q_\la$ is a homogeneous polynomial of degree $|\la|$, at $\textbf{t}=0$,
 \[ \tau_{N}(0) = Q_{\emptyset} =1. \]
  Hence $\tau_N$ and
 $\tau_{BGW}^{(N)}$ satisfy the same Virasoro constraints with the same normalization condition. By Alexandrov's
 theorem in \cite{Alex18},   we have
\[\tau_{BGW}^{(N)}=\tau_{N} \]
for all $N$. This completes the proof of Theorem \ref{thm:A=gBGW}.

By equation \eqref{eqn:Q/Q}, Theorem \ref{thm:A=BGW} follows from Theorem \ref{thm:A=gBGW} with $N=0$.
Note that the dimension constraint for the geometric interpretation of $\tau_{BGW}$ corresponds to the
fact that the coefficient of $\hbar^m$ in $\tau_N$ is a homogeneous polynomial of $(t_1, t_3, \cdots)$  of degree $m$
for all $m$.
$\Box$

\appendix
\vspace{30pt}
\hspace{160pt} {\bf \Large Appendix}

\section{Some elementary identities}
\begin{lem}    \label{lem:rho} 
	For any three positive integers $a<b<c$, and three non negative integers $\mu_a, \mu_b, \mu_c$ labeled by $a,b,c$, we have
	\begin{align*}
	\sum_{(i,j,m) \subseteq P(a,b,c)} (-1)^{i+\tilde{j}(i)+\tilde{m}(j,i)}\rho_k(\mu_i,\mu_j,\mu_m)=0
	\end{align*}
	for $k=0, 1, 2$, where $P(a,b,c)$ is the set of all permutations of $(a,b,c)$,
    $\tilde{j}(i)$ and $\tilde{m}(i,j)$ are defined by equation \eqref{eqn:tilde},
    $\rho_k$ are defined by equation \eqref{eqn:defrho}.
\end{lem}
{\bf Proof:}
Set
\[K_{abc} :=(\mu_a+1)(\mu_b+1)(\mu_c+1).\]
Note that
\begin{align*}
\rho_0(\mu_i,\mu_j,\mu_m)-\rho_0(\mu_i,\mu_m,\mu_j)
        =&  \left\{ -16 K_{abc}+\theta(1)^2 \right\} \cdot \mu_i(\mu_j-\mu_m), \\
\rho_1(\mu_i,\mu_j,\mu_m)-\rho_1(\mu_i,\mu_m,\mu_j)
        =&\left\{16 K_{abc}(1-\mu_i)-\theta(1)^2 \right\} \cdot(\mu_j-\mu_m), \\
\rho_2(\mu_i,\mu_j,\mu_m)-\rho_2(\mu_i,\mu_m,\mu_j)
        =& K_{abc}\cdot(\mu_j-\mu_m).
\end{align*}
The lemma then follows from the following identities
\begin{align*}
&\mu_i(\mu_j-\mu_m)+\mu_j(\mu_m-\mu_i)+\mu_m(\mu_i-\mu_j)=0 ,\\
& (\mu_j-\mu_m)+(\mu_m-\mu_i)+(\mu_i-\mu_j)=0.
\end{align*}
$\Box$

\begin{lem}\label{lem:3identities}
For any three positive integers $a<b<c$, and three non negative integers $\mu_a, \mu_b, \mu_c$ labeled by $a,b,c$, we have
\begin{align*}
\sum_{(i,j,k)\in\sigma(a,b,c)}(\mu_j-\mu_m)=
\sum_{(i,j,k)\in\sigma(a,b,c)}\mu_i(\mu_j-\mu_m)=
\sum_{(i,j,k)\in\sigma(a,b,c)}(\mu_j\mu_m-\mu_i^2)(\mu_j-\mu_m)=0,
\end{align*}
where $\sigma(a,b,c)=\{(a,b,c),(b,c,a),(c,a,b)\}$.
\end{lem}
The proof of this lemma is straightforward.


\vspace{30pt} \noindent
Xiaobo Liu \\
School of Mathematical Sciences \& \\
Beijing International Center for Mathematical Research, \\
Peking University, Beijing, China. \\
Email: {\it xbliu@math.pku.edu.cn}
\ \\ \ \\
Chenglang Yang \\
School of Mathematical Sciences \& \\
Beijing International Center for Mathematical Research, \\
Peking University, Beijing, China. \\
Email: {\it yangcl@pku.edu.cn}

\end{document}